\documentclass[12pt,a4paper,twoside]{article}
\usepackage[top=0.5in, bottom=1in, left=0.5in, right=0.5in]{geometry}
\usepackage[pdftex]{graphicx}
\usepackage{listings}
\usepackage{xcolor}

\usepackage{color}
\usepackage{lscape}
\usepackage{epstopdf}
\usepackage{amsmath}
\usepackage{hyperref}
\usepackage{setspace}
\usepackage{listings}
\usepackage{pdfpages}
\usepackage{rotating} 

\usepackage{amsmath}
\usepackage{array} 
\usepackage{natbib}
\usepackage[utf8]{inputenc}

\usepackage{amssymb} 
\usepackage{titlesec}
\usepackage{acronym}
\usepackage{hyperref}
\usepackage{fmtcount}
\usepackage{etoolbox}
\usepackage{tocloft}
\newcommand{\listappendicesname}{}
\newlistof{appendices}{apc}{\listappendicesname}

\lstdefinestyle{pythonstyle}{
    language=Python,
    basicstyle=\ttfamily\footnotesize,
    keywordstyle=\color{blue},
    stringstyle=\color{green!60!black},
    commentstyle=\color{gray},
    numbers=none,
    breaklines=true,
    showstringspaces=false,
    tabsize=4
}

\newenvironment{changemargin}[2]{%
\begin{list}{}{%
    \setlength{\leftmargin}{#1}%
    \setlength{\rightmargin}{#2}%
}%
\item[]}
{\end{list}}

\begin{document}
\baselineskip=0.27in
{\bf \LARGE
\begin{changemargin}{-0.5cm}{-0.5cm}
\begin{center}
{Quantum Fisher Information in Curved Spacetime: Dirac Particles in Noisy Channels around a Schwarzschild Black Hole}
\end{center}\end{changemargin}}
\vspace{4mm}
\begin{center}
\large{Cookey Iyen $^{a,}$}\footnote{\scriptsize E-mail:~ iyen@fuwukari.edu.ng;}, \large{Muhammad Sanusi Liman ${}^{b}$}, \large{Benedict O. Ayomanor ${}^{c}$}, \large{Emem-obong Solomon James${}^{b}$}, \large{Yame Mwanzang Philemon $^{b}$} and \large{Babatunde James Falaye $^{b,d,}$}\footnote{\scriptsize E-mail:~ bfalaye@aul.edu.ng;~babatunde.falaye@science.fulafia.edu.ng}
\end{center}
{\footnotesize
\begin{center}
{\it $^\textbf{a}$Department of Pure and Applied Physics, Federal University Wukari, Nigeria.}
{\it $^\textbf{b}$Department of Physics, Federal University of Lafia,  P. M. B. 146, Lafia, Nigeria.}
{\it $^\textbf{c}$Department of Science Laboratory Technology, Federal University of Petroleum Resources, Nigeria}
{\it $^\textbf{d}$Department of Physics, Anchor University, Lagos, Nigeria.}
\end{center}
}
\begin{abstract}
\noindent
Quantum information processing promises significant advantages over classical methods but remains vulnerable to decoherence induced by environmental interactions and spacetime effects. This work investigates the behavior of Quantum Fisher Information (QFI) as a diagnostic tool for entanglement and parameter estimation in a three-qubit entangled Dirac system subjected to dissipative noisy channels in the curved spacetime of a Schwarzschild black hole. In particular, we examine the influence of the squeezed generalized amplitude damping (SGAD) channel, along with its subchannels -- generalized amplitude damping (GAD) and amplitude damping (AD) -- on the QFI with respect to entanglement weight ($\theta$) and phase ($\phi$) parameters. Our results show that under strong squeezing ($r = 1$), the QFI with respect to $\theta$ becomes completely resistant to variations in the Hawking temperature ($T_H$), while still exhibiting degradation with increasing channel temperature ($T_C$). The QFI decay is significantly slower at $r = 1$ compared to $r = 0$, suggesting that squeezing can function as an error mitigation strategy. For QFI with respect to $\phi$, a transient spike is observed at $T_C = 2$, potentially due to thermal resonance or non-monotonic decoherence, and this behavior is unaffected by $T_H$. Similar patterns are noted in the GAD and AD channels, where $T_C$ consistently dominates as the principal source of decoherence. Overall, the results highlight the intricate interplay between environmental noise, relativistic effects, and quantum error resilience in curved spacetime.
\\
\\
Keywords: Quantum Fisher Information, Dirac System, Squeezed Generalized Amplitude Damping Channel, Hawking Radiation, Schwarzschild Black Hole Background
\end{abstract}


\section{Introduction}
Quantum information and computation is the study of information processing tasks that use quantum mechanical systems that utilise the state of a quantum system to encode information \citep{krasnoholovets2022information}. The central unit of quantum information is the quantum bit, or qubit, while Quantum communication is defined as a secure and unconditional communication between two or more participants that adheres to the laws of quantum physics \citep{mishra2022key,azahari2023quantum}, and it involves teleporting messages through quantum carriers or tunnels. Quantum communication is made possible by a unique quantum resource called quantum entanglement (QE). QE is a quantum effect in which particles originating from the exact same source exhibit a linked existence, irrespective of their physical separation. Any alteration performed on one of the entangled particles will have an impact on the quantum state of other particles they are entangled with \citep{iyen2023scrutinizing,einstein1935can}. 

Quantum communication offers many advantages, such as unconditional security, no cloning theorem protection \citep{zygelman2024no,venkatesh2024lightweight}, eavesdropping detection \citep{castro2022enhancing,lee2022eavesdropping}, improved communication efficiency, and enhanced privacy in data communication. However, it also has several challenges. These challenges encompass concerns regarding scalability \citep{bakhshinezhad2024scalable,uppu2021quantum}, decoherence and noise \citep{khan2024quantum,martinez2022decoherence}, lack of quantum repeaters, hardware and infrastructure, among others. This research is concerned with the decoherence challenge, which is the loss of entanglement in an entangled quantum system. Decoherence may be attributed to the interaction between a quantum system and its environment. Since quantum communication is based on QE, it implies that whatever affects QE also affects quantum communication. As a matter of fact, decoherence is linked to the loss of information in quantum communication protocols.

Quantum metrology is the study of making high-resolution and highly sensitive measurements of classical and quantum parameters using quantum theory \citep{hu2024quantum}. It is concerned with parameter estimation in quantum mechanics; it may also be defined as the study of precision measurement of an unknown parameter \citep{zhou2025randomized}. There are a number of quantum metrology techniques, such as quantum interferometry \citep{jin2024quantum,marshall2022high}, squeezed state metrology \citep{sinatra2022spin,li2023multi}, entanglement-enhanced metrology \citep{huang2024entanglement,long2022entanglement,deng2024quantum}, QFI \citep{altherr2021quantum,rath2021quantum,gorecki2022quantum}, and others. Since whatever affects QE affects quantum communication, the approach of this research is to use the QFI quantum metrology techniques to estimate the entanglement parameter before and after a quantum system encounters noisy channels. Thereby estimating the effect of the noisy channel on QE and quantum communication. 

There are different noisy channels in quantum information, such as bit flip \citep{das2025investigating}, phase flip \citep{hu2020protecting}, GAD \citep{khatri2020information}, depolarising channel \citep{zlotnick2025entanglement}, dephasing channel \citep{rahman2022fidelity}, AD channel \citep{chessa2021quantum}, the SGAD \citep{iyen2024examining}, and lots more. In this research we consider the effect of the SGAD noise channel on quantum entanglement. It is necessary, when investigating the behaviour of realistic quantum systems, to account for environmental disturbance in addition to spacetime curvature, gravity, and other cosmic characteristics. In this research endeavour, the Schwarzschild black hole spacetime notion is utilised for this purpose. A black hole can be defined as a spatial region characterised by an exceptionally dense matter core, rendering any adjacent object incapable of escaping its gravitational attraction \citep{han2024mass,kunstatter2022general,zhao2023trapped}. 

The event horizon, the boundary surrounding the black hole from which nothing can be retrieved, and the singularity, the region at the black hole's interior with an infinite density and nil volume, where the physical laws are violated, are two characteristics that define black holes. It is widely believed that black holes exert an extraordinarily strong gravitational pull by virtue of their enormous mass. Based on rotation and charge, there are 4 types of black holes, namely, the Schwarzschild Black Hole \citep{wu2024genuinely,alonso2022effective}, the Kerr Black Hole \citep{horowitz2023extremal,cangemi2023kerr}, the Reissner-Nordstr\"{o}m Black Hole \citep{senjaya2024exact,shaymatov2021charged}, and the Kerr-Newman Black Hole \citep{foo2021hawking,sun2021entanglement}. In this research work, we consider the Schwarzschild black hole, as it provides the simplest solution to Einstein's field equation, and also it emits the Hawking radiation, which can lead to decoherence and loss of information in quantum communication. 

Many researchers have worked on the effect of noisy channels on quantum communications; some have investigated the effect of noisy channels on quantum communication protocols such as quantum teleportation (QT), quantum key distribution (QKD), and others \citep{harraz2021protected, zidan2023quantum,mafu2022security,shu2023quantum,falaye2017investigating,adepoju2017joint}. For example, \citep{iyen2023scrutinizing} investigated the effect of dissipative noisy channels, namely the SGAD channel, on the Joint Remote State Preparation (JRSP) quantum communication protocol. They found out that though the fidelity of the JRSP quantum communication protocol degraded with increasing temperature, it is possible to enhance the fidelity by varying the squeezing parameter $r$ and $\Phi$ while \citep{mafu2022security} investigated how the collective rotation noise affects the security of the BB84 QKD protocol; they found that the BB84 protocol is robust against intercept-resend attacks on collective-rotation noise channels when the rotation angle is varied arbitrarily within particular bounds. Others have investigated the effect of noisy channels on the quantum entanglement parameter of the entangled quantum system \citep{iyen2024examining,oh2021quantum,gorecki2022quantum,zhai2023control}. 

For example, Giovanni {\it et al.}, \citep{ragazzi2024generalized} examined metrological scenarios where the parameter of interest is encoded onto a quantum state through the action of a noisy quantum gate and investigated the ultimate bound to precision by analysing the behaviour of the QFI. By examining the interplay of various gate imperfections, they formulated a model of generic noise and discovered intriguing scenarios where the coexistence of two distinct noise types results in a more robust QFI or, surprisingly, even enhances the QFI compared to when only a single type of noise is present, Falaye and Liman \citep{falaye2020probing} looked into how the Amplitude Damping (AD) and Phase Damping (PD) channels affect the QFI of open Dirac systems influenced by Hawking effects from a Schwarzschild black hole. They found that the QFI was more attenuated with decoherence in the presence of the Hawking radiation from the Schwarzschild black hole. Iyen {\it et al.} \citep{iyen2024examining} investigated the QFI of entangled Dirac particles in an SGAD channel; however, the present study advances their work by examining the impact of Hawking emission from a Schwarzschild black hole on the QFI of such particles interacting with an SGAD channel.

This research paper is arranged in the following order: In section \ref{one}, we introduce the concept of the QFI. In section \ref{two} we investigate the QFI of Dirac particles in contact with a dissipative noise channel and in the surroundings of a Schwarzschild BH. In section \ref{three}, we discuss our methodology. We present our research outcomes in section \ref{four}, while discussion of the results and conclusions is made in section \ref{five}.
    
\section{Quantum Fisher Information}\label{one}
QFI is a notion within quantum metrology, the field focused on achieving high-precision measurements through quantum systems. It measures the amount of information embedded in a specific parameter, like the phase of a quantum wavefunction. Conversely, it assesses how sensitive a quantum state is to changes in a particular variable.

The core concept of QFI is that a quantum state $\rho_\lambda$ depends on an undisclosed parameter $\lambda$. The QFI can be represented using the spectrum decomposition $\rho_\lambda=\sum_{i=1}^Np_i|\psi_i\rangle\langle\psi_i|$, where $\{|\psi_i\rangle\}$ constitutes a complete and orthogonal basis, and $p_i$ denotes the weight assigned to $|\psi_i\rangle$:
\begin{equation}\label{062811}
F_\lambda=\underbrace{\sum_{i=1}^M\frac{1}{p_i}\left(\frac{\partial pi}{\partial \lambda}\right)^2}_{(I)} + 
\underbrace{\sum_{i=1}^M p_iF_{\lambda,i}}_{(II)} -\underbrace{\sum_{i\neq j}^M\frac{8p_ip_j}{p_i + p_j}\biggl\lvert\langle\psi_i\lvert\frac{\partial \psi_i}{\partial \lambda}\biggr\rangle\biggr \lvert^2}_{(III)},
\end{equation}
where the expression $F_{\lambda,i}$ corresponds to:
\begin{equation}\label{062812}
F_{\lambda,i}=4\left(\left\langle\frac{\partial\psi_i}{\partial\lambda}\right\rvert\frac{\partial\psi_i}{\partial \lambda}\right\rangle -\biggr\rvert\langle\psi_i\rvert\frac{\partial\psi_i}{\partial\lambda}\biggr\rangle\biggr\lvert\biggr).
\end{equation}
In Eq. (\ref{062811}), (I) represents the classical Fisher information of a probability distribution. The QFI of the pure state is denoted as (II). As (III) is a composite of pure states, it leads to a reduction in the total QFI. 

\section{The Kruskal Vacuum near a Schwarzschild Black Hole}
The Schwarzschild space-time metric may be expressed as:
\begin{equation}\label{06281}
ds^2=-(1-\frac{2M}{r})dt^2 + (1-\frac{2M}{r})^{-1}dr^2 +r^2(d\theta^2+sin^2\theta d\varphi^2),
\end{equation}
In Eq. (\ref{06281}), M stands for the mass of the BH. To simplify the equation, the parameters G, c, $\hbar$, and $k_B$ are assumed to have a value of 1. The Dirac equation that corresponds to the Schwarzschild BH is represented as Eq. (\ref{06282}).
\begin{equation}\label{06282}
[\gamma^ae_a^\mu(\partial_\mu + \Gamma_\mu)]\Psi=0.
\end{equation}
By using equations (\ref{06281}) and (\ref{06282}), we derive the Dirac equation for the Schwarzschild space-time, as shown in Eq. (\ref{06283}).

\begin{equation}\label{06283}
-\frac{\gamma_0}{\sqrt{1-\frac{2M}{r}}}\frac{\partial\Psi}{\partial t}+\gamma_1\sqrt{1-\frac{2M}{r}}[\frac{\partial}{\partial r} + \frac{1}{r}+\frac{M}{2r(r-2M)}]\Psi +\frac{\gamma_2}{r}(\frac{\partial}{\partial\theta}+\frac{1}{2}cot\theta)\Psi + \frac{\gamma_3}{rsin\theta}\frac{\partial\Psi}{\partial\varphi}=0.
\end{equation}
By using the following wavefunction alternative as a parametric substitution:
\begin{equation}\label{06284}
\Psi=(1-\frac{2M}{r})^{-\frac{1}{2}}
\begin{pmatrix}
\frac{i\zeta_1^\pm(r)}{r}\phi_{jm}^{\pm}(\theta,\varphi) \\
\frac{i\zeta_2^\mp(r)}{r}\phi_{jm}^{\mp}(\theta,\varphi)
\end{pmatrix} e^{-i\omega t},
\end{equation}
where the angular harmonics of the spinor are :
\begin{equation}\label{06285}
\phi^+_{jm}=\begin{pmatrix}
\sqrt{\frac{j+m}{2j}}Y_l^{m-\frac{1}{2}}\\
\sqrt{\frac{j-m}{2j}}Y_l^{m+\frac{1}{2}}
\end{pmatrix}, for j= l+\frac{1}{2},
\end{equation}
\begin{equation}\label{06286}
\phi^-_{jm}=\begin{pmatrix}
\sqrt{\frac{j+1-m}{2j+2}}Y_l^{m-\frac{1}{2}}\\
-\sqrt{\frac{j+1+m}{2j+2}}Y_l^{m+\frac{1}{2}}
\end{pmatrix}, for j= l-\frac{1}{2}.
\end{equation}
By solving Eq. (\ref{06283}), we may get the positive frequency outgoing solutions for both the interior and exterior parts of the event horizon as:

\begin{equation}\label{06287}
\Psi_k^{II+}=\mathcal{F}e^{i\omega\mu} (r<r_+)
\Psi_k^{I+}=\mathcal{F}e^{-i\omega\mu} (r>r_+),
\end{equation}

The 4-component Dirac spinor $\mathcal{F}$ is defined by:
\begin{equation}\label{06288}
\mathcal{F}=
\begin{bmatrix}
i(r^4-2Mr^3)^{-\frac{1}{4}}\phi^{\pm_jm}(\theta,\varphi)\\
(r^4-2Mr^3)^{-\frac{1}{4}}\phi^{\mp_jm}(\theta,\varphi)
\end{bmatrix} e^{-i\omega t},
\end{equation}
In Eq. (\ref{06287}), the variable $\mu$ is defined as the dissimilarity between $t$ and $r^*$. The value of $r^*$ is obtained by adding $r$ to the product of $2M$ and the natural logarithm of the fraction $\frac{(r-2M)}{(2M)}$, which represents the tortoise coordinates, whereas $\omega$ is the monochromatic frequency associated with the Dirac field. The modes are characterised by the wavevector k. The $\Psi_k^{II+}$ and $\Psi_k^{I+}$ constitute a full orthogonal family, which is due to the fact that they are analytic both within and on the exterior of the event horizon. After Domour-Ruffni's recommendation \citep{damour1976black}, to provide a comprehensive basis for positive modes of energy, we construct an analytical continuation for Eq. (\ref{06287}). The Bogoliubov transformations are obtained by quantising the Dirac field in Schwarzschild and Kruskal spacetime, as shown in Eq. (\ref{06289}).

\begin{equation}\label{06289}
\begin{bmatrix}
c_k^\sigma\\
d^{\sigma\dagger}_{-k}
\end{bmatrix}
=\mathcal{V}
\begin{bmatrix}
a_k^\sigma\\
b^{\sigma\dagger}_{-k}
\end{bmatrix}
\mathcal{V}^{-1}.
\end{equation}
The variable $\mathcal{V}=exp [r(a^{I\dagger}_k b^{II\dagger}_{-k}+ a^I_k b^{II}_{-k})]$ is used to represent a two-mode Dirac squeezing operator. The annihilation and creation operators of the Kruskal vacuum are denoted as $c_k$ and $d_k$, respectively. The Kruskal vacuum for mode k is obtained by sufficiently normalising the state vector, as shown below:
\begin{equation}\label{062810}
\begin{cases}
|0\rangle_K^+\rightarrow (e^{-\frac{\omega}{T}+1})^{-\frac{1}{2}}|0\rangle_I^+|0\rangle_{II}^- + (e^{\frac{\omega}{T}+1})^{-\frac{1}{2}}|1\rangle_I^+|1\rangle_{II}^-,\\
|1\rangle_K^+ \rightarrow |1\rangle_I^+|0\rangle^-_{II},
\end{cases}
\end{equation}

where $\{|n_k\rangle_I^+\}$ and $\{|n_k\rangle_{II}^-\}$ represent the orthonormal bases for the areas within and on the exterior of the event horizon, respectively. $T=(8\pi M)^{-1}$ represents the Hawking temperature. The + and – superscripts on the kets stand for the vacua, representing both the particle and the antiparticle.

\section{Methodology}\label{three}
\begin{figure}[!t]
\centering
\includegraphics[scale=0.5]{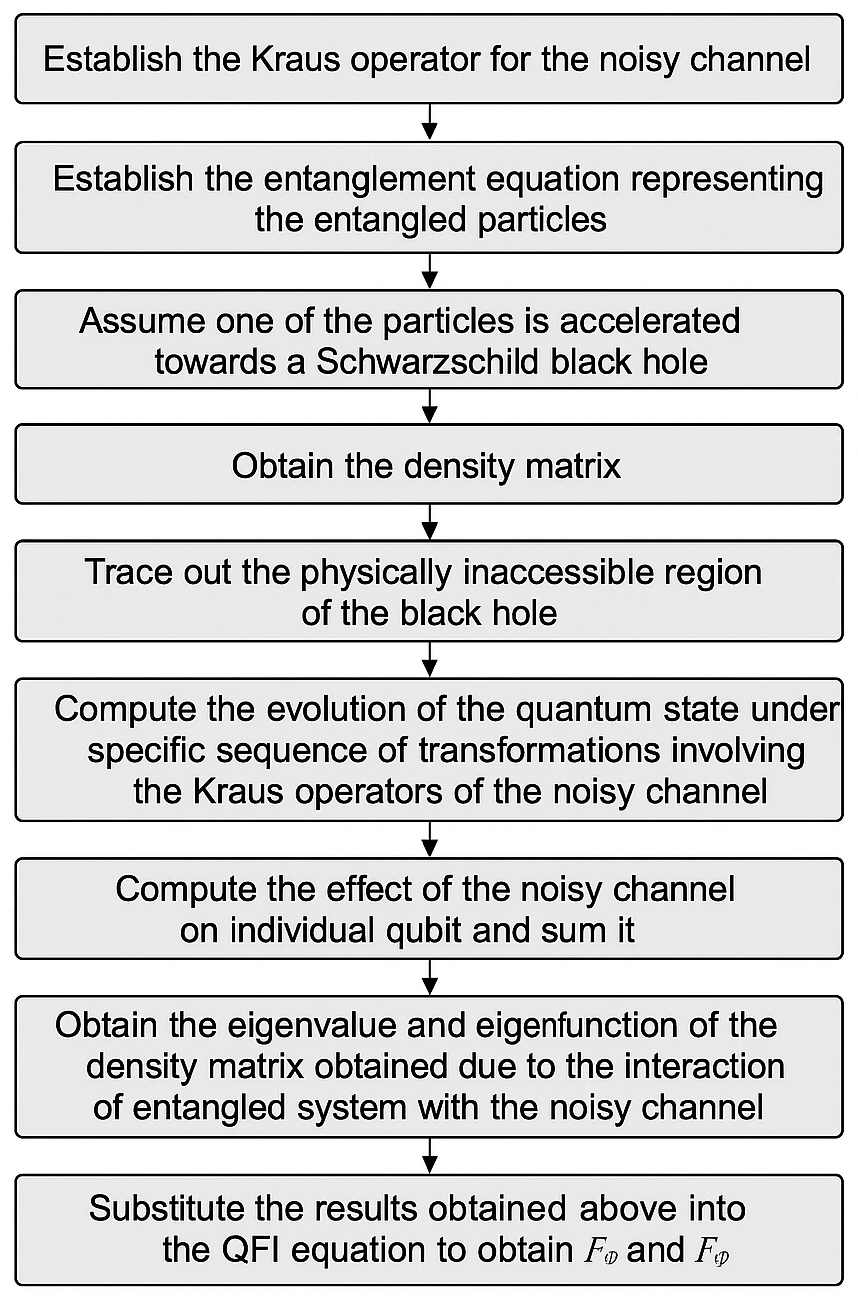}
\caption{The methodology for the research}
\label{fig3103a}
\end{figure}

The methodological framework used in this research is visually summarized in Figure~\ref{fig3103a}. It describes the sequential steps of the investigation into the behavior of QFI in the curved spacetime of a Schwarzschild black hole.

As seen in Figure~\ref{fig3103a}, the study begins with the theoretical formulation of the Dirac equation in Schwarzschild spacetime. This establishes the relativistic foundation for analyzing entangled Dirac particles. Next, the Kruskal vacuum is used to accurately simulate the quantum field near the event horizon, allowing for the introduction of Hawking radiation effects via Bogoliubov transformations.

A three-qubit entangled state is then initialized in Minkowski space, simulating a quantum communication setup with three observers: Alice, Bob, and Caleb. The arrangement, as shown in Figure~\ref{fig:1113c}, proposes that Alice and Bob are exposed to a dissipative noisy channel (SGAD), whereas Caleb's qubit interacts with Hawking radiation due to acceleration into a Schwarzschild black hole.

The evolution of this tripartite system is investigated utilizing quantum channels, namely the SGAD channel and its subchannels, GAD and AD. Each noise model is mathematically defined by a collection of Kraus operators, which are employed to compute the system's density matrix during decoherence.

The resulting density matrices' eigenvalues and eigenvectors are then retrieved and inserted into the QFI formula (Eq.~\ref{062811}). This allows for evaluation of QFI based on two key parameters: entanglement weight $\theta$ and phase $\phi$. Finally, the QFI is calculated and shown as a function of environmental factors (channel temperature $T_C$) and relativistic parameters (Hawking temperature $T_H$), with the goal of understanding how squeezing and gravitational effects interact with quantum coherence.

This step-by-step approach ensures a thorough investigation of how relativistic and ambient noise affect quantum information contained in entangled Dirac systems.
\section{Results}\label{four}
\subsection{QFI in Dissipative Noisy Channel in the environment of a Schwarzschild BH}\label{two}
This section explores the characteristics of QFI in a noisy channel affected by a dissipative noisy channel, which is the SGAD channel and its associated subchannels, namely the AD and GAD channels within the environment of a Schwarzschild BH.

\subsubsection{The squeezed generalized amplitude damping channel}
The SGAD channel is a quantum noise model that generalises the standard AD channel by including the attributes of both thermal noise and squeezing. It describes the effects of the contact of a quantum setup with a squeezed thermal reservoir and is relevant in quantum optics and quantum information processing, particularly in quantum communication protocols.

Mathematically, the SGAD channel is described by a set of Kraus operators that govern the evolution of a single-qubit density matrix under the influence of a squeezed thermal bath. The Kraus operators for the SGAD channel are given by\citep{ghosh2012surface, Nielsen20001}:
\begin{equation}\label{q6}
\begin{split}
E_0^{S}=&
\sqrt{Q}
	\begin{bmatrix} 
	1 & 0  \\
	0 & \sqrt{1-\lambda}
	\end{bmatrix},\\
	\quad
E_1^{S}=&
\sqrt{Q}
	\begin{bmatrix} 
	0 & \sqrt{\lambda}  \\
	0 & 0
	\end{bmatrix},\\
	\quad
E_2^{S}=&
	\sqrt{1-Q}
	\begin{bmatrix} 
	\sqrt{1-v} & 0  \\
	0 & \sqrt{1-\mu}
	\end{bmatrix},\\
	\quad
E_3^{S}=&
	\sqrt{1-Q}
	\begin{bmatrix} 
	0 & \sqrt{\mu}e^{i\Phi}  \\
	\sqrt{v} & 0 \\
	\end{bmatrix},
	\quad
	\end{split}
\end{equation}
In Eq. (\ref{q6}), the parameters $\mu$, v, and $\lambda$ are defined by the following equations:
\begin{equation}\label{q6a}
\begin{split}
\mu=\frac{2N+1}{2N(1-Q)} \frac{\sinh^2(\frac{\gamma_0 a}{2})}{\sinh^2 (\gamma_0 \frac{2N+1}{2})} e^{\frac{-\gamma_0 (2N+1)}{2}},\\
v=\frac{N}{(1-Q)(2N+1)}(1-e^{-\gamma_0 (2N+1)}),\\
\lambda=\frac{1}{Q}(1-(1-Q)(\mu+v)-e^{(-\gamma_0(2N+1))},
\end{split}
\end{equation}
In equations (\ref{q6a}), the value of $a$ is determined by the expression $a=\sinh(2r)(2N_{th}+1)$. The value of $N$ is given by $N=N_{th}(\cosh^2(r)+\sinh^2(r))+Sinh^2(r)$. Here, $N_{th}$ is defined as $\frac{1}{e^{(\frac{\hbar \omega}{k_B T})}-1}$. It is worth noting that the time parameter has been disregarded in these equations.

The SGAD channel plays a significant role in studying the impact of quantum noise on entanglement, coherence, and other quantum resources, particularly in the presence of nonclassical environmental effects.
\subsubsection{QFI in SGAD channel in the background of a Schwarzschild Black Hole}
assuming a flat Minkowski space-time with Alice, Bob, and Caleb as the three individuals connected by a 3-qubit entanglement state beginning at the same place.  Alice possesses a detector that can identify mode $|n\rangle_a$ when decoherence is present. On the other hand, as shown in Figure (\ref{fig:1113c}), Bob and Caleb use detectors that respond to modes $|n\rangle_b$ and $|n\rangle_c$, respectively. We suppose that a dissipative noisy channel affects the entanglement between Alice and Bob and that the Hawking rays from a Schwarzschild BH affect the entanglement on the Caleb side. Let the initial entanglement between Alice, Bob and Caleb be given by:

\begin{equation}\label{0831001}
|\chi\rangle_{ABC}=cos(\theta)|0\rangle_A|0\rangle_B|0\rangle_C+sin(\theta)e^{i\phi}|1\rangle_A|1\rangle_B|1\rangle_C,
\end{equation}
where the weight and phase parameters are denoted by $\theta$ and $\phi$, respectively. The QFI in the SGAD channel in the setting of a Schwarzschild BH is thoroughly investigated using the Eqs.  (\ref{0831001}) and (\ref{062810}). The Eq. (\ref{0831001}) can be depicted in relation to the BH perspective for Caleb and the Minkowski perspective for both Alice and Bob. After making applicable replacements and substitutions, we have:
\begin{figure}[!t]
	\centering
		\includegraphics[width=1.00\textwidth]{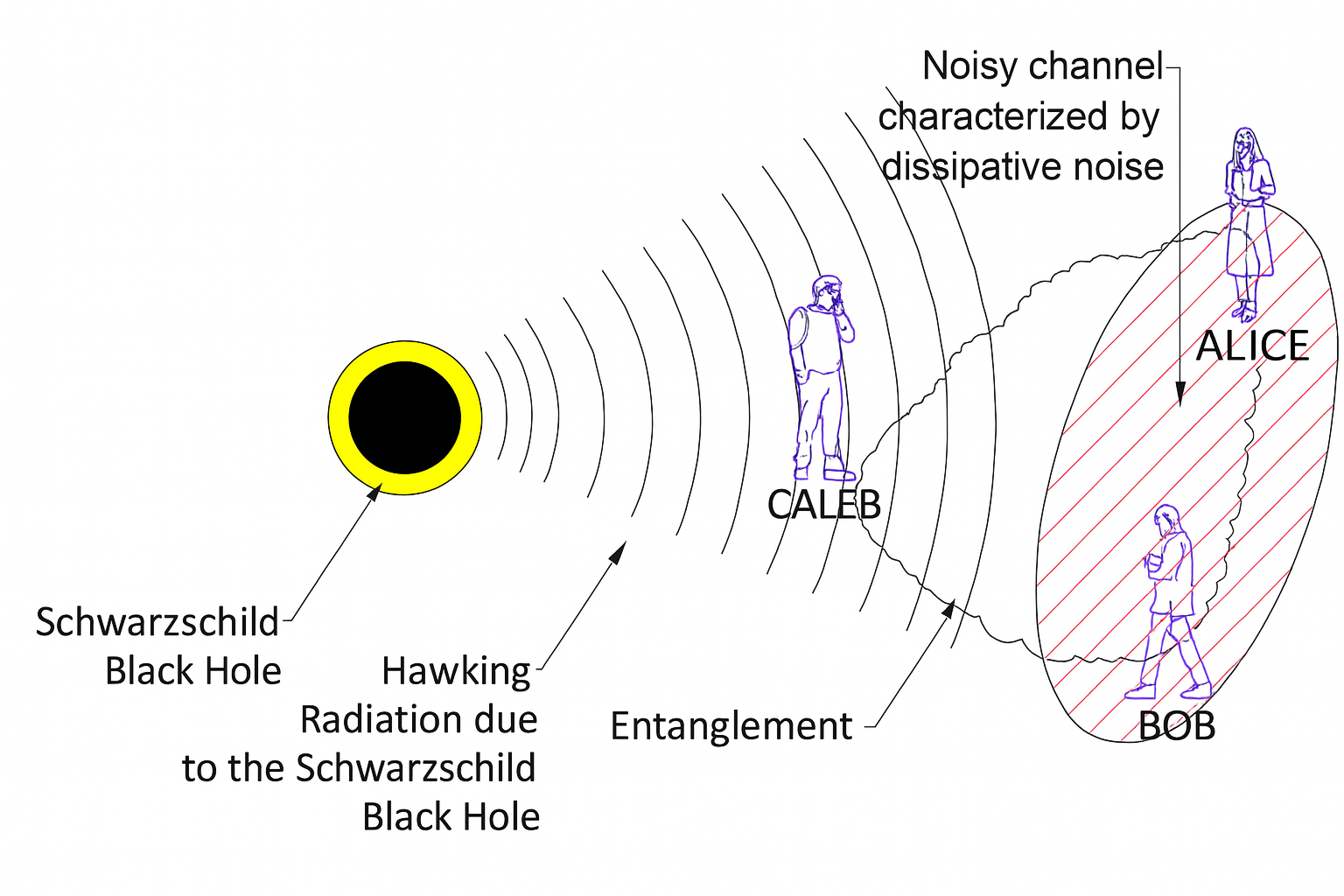}
	\caption{Illustration of a three-qubit entangled system in which the third qubit is subjected to Hawking radiation from a Schwarzschild black hole, while the first and second qubits experience the effects of a dissipative noisy channel.}
\label{fig:1113c}
\end{figure}
\begin{equation}\label{101101}
|\chi^{\prime} \rangle =A\left|0_{\hat{a}},0_{\hat{b}},0_{\hat{\text{c1}}},0_{\hat{\text{c2}}}\right\rangle +B\left|1_{\hat{a}},1_{\hat{b}},1_{\hat{\text{c1}}},0_{\hat{\text{c2}}}\right\rangle +F\left|0_{\hat{a}},0_{\hat{b}},1_{\hat{\text{c1}}},1_{\hat{\text{c2}}}\right\rangle,
\end{equation}
where $A= \sin\theta e^{i\phi}$, $B=\cos\theta(e^{-{\frac{\omega}{T}}}+1)^{-\frac{1}{2}}$, and $F=\cos\theta (e^{\frac{\omega}{T}}+1)^{-\frac{1}{2}}$. By tracing over the state of the black hole's inner region, which is causally isolated from its outer region,  we obtain:
\begin{equation}\label{101102}
\begin{split}
\rho_{abc1} =& A^2 \left|0_{\hat{a}},0_{\hat{b}},0_{\hat{\text{c1}}}\right\rangle \cdot \left\langle 0_{\hat{a}},0_{\hat{b}},0_{\hat{\text{c1}}}\right|+A B \left|0_{\hat{a}},0_{\hat{b}},0_{\hat{\text{c1}}}\right\rangle \cdot \left\langle 1_{\hat{a}},1_{\hat{b}},1_{\hat{\text{c1}}}\right|\\
+&A B \left|1_{\hat{a}},1_{\hat{b}},1_{\hat{\text{c1}}}\right\rangle \cdot \left\langle 0_{\hat{a}},0_{\hat{b}},0_{\hat{\text{c1}}}\right|+B^2 \left|1_{\hat{a}},1_{\hat{b}},1_{\hat{\text{c1}}}\right\rangle \cdot \left\langle 1_{\hat{a}},1_{\hat{b}},1_{\hat{\text{c1}}}\right|\\
+&F^2 \left|0_{\hat{a}},0_{\hat{b}},1_{\hat{\text{c1}}}\right\rangle \cdot \left\langle 0_{\hat{a}},0_{\hat{b}},1_{\hat{\text{c1}}}\right|,
\end{split}
\end{equation}
which enables us to ascertain the authentic, accessible density matrix that connects Alice, Bob, and Caleb. As Caleb accelerates towards a Schwarzschild black hole background, the density matrix undergoes a transformation which is gotten by making substitutions from Eq. (\ref{062810}) and takes the form described by Eq. (\ref{101103}).
\begin{equation}\label{101103}
\begin{split}
\mathcal{K}_{SGAD}\rho_{abc} =& A B Q^2 \bar{\lambda } \left(\left|0_{\hat{a}},0_{\hat{b}},0_{\hat{\text{c1}}}\right\rangle \right)\cdot \left\langle 1_{\hat{a}},1_{\hat{b}},1_{\hat{\text{c1}}}\right|+A B Q^2 \bar{\lambda } \left(\left|1_{\hat{a}},1_{\hat{b}},1_{\hat{\text{c1}}}\right\rangle \right)\cdot \left\langle 0_{\hat{a}},0_{\hat{b}},0_{\hat{\text{c1}}}\right|\\
+&B^2 Q^2 \bar{\lambda }^2 \left(\left|1_{\hat{a}},1_{\hat{b}},1_{\hat{\text{c1}}}\right\rangle \right)\cdot \left\langle 1_{\hat{a}},1_{\hat{b}},1_{\hat{\text{c1}}}\right|\\
+&B^2 \lambda  \mu  Q e^{2 \text{i$\Phi $}} \bar{Q} \left(\left|0_{\hat{a}},0_{\hat{b}},1_{\hat{\text{c1}}}\right\rangle \right)\cdot \left\langle 0_{\hat{a}},0_{\hat{b}},1_{\hat{\text{c1}}}\right|+B^2 \lambda  Q \bar{\mu } \bar{Q} \left(\left|0_{\hat{a}},1_{\hat{b}},1_{\hat{\text{c1}}}\right\rangle \right)\cdot \left\langle 0_{\hat{a}},1_{\hat{b}},1_{\hat{\text{c1}}}\right|\\
+&A^2 Q^2 \left(\left|0_{\hat{a}},0_{\hat{b}},0_{\hat{\text{c1}}}\right\rangle \right)\cdot \left\langle 0_{\hat{a}},0_{\hat{b}},0_{\hat{\text{c1}}}\right|+B^2 \lambda ^2 Q^2 \left(\left|0_{\hat{a}},0_{\hat{b}},1_{\hat{\text{c1}}}\right\rangle \right)\cdot \left\langle 0_{\hat{a}},0_{\hat{b}},1_{\hat{\text{c1}}}\right|\\
+&F^2 Q^2 \left(\left|0_{\hat{a}},0_{\hat{b}},1_{\hat{\text{c1}}}\right\rangle \right)\cdot \left\langle 0_{\hat{a}},0_{\hat{b}},1_{\hat{\text{c1}}}\right|.
\end{split}
\end{equation}
The eigenvalues of $\mathcal{K}_{SGAD}\rho_{abc}$ are obtained as shown below:
\begin{equation}\label{101201}
\left\{0,0,0,0,0,Q \left(B^2 \lambda  \mu  e^{2 \text{i$\Phi $}} \bar{Q}+Q \left(B^2 \lambda ^2+F^2\right)\right),Q^2 \left(B^2 \bar{\lambda }^2+A^2\right),B^2 \lambda  Q \bar{\mu } \bar{Q}\right\}.
\end{equation}
From Eq. (\ref{101201}), the eigenvalues of $\mathcal{K}_{SGAD}\rho_{abc}$ are extracted as \\
$|\text{e1}\rangle =Q \left(B^2 \lambda  \mu  e^{2{i\Phi }} \bar{Q}+Q \left(B^2 \lambda ^2+F^2\right)\right)$, 
$|\text{e2}\rangle =Q^2 \left(B^2 \bar{\lambda }^2+A^2\right)$,
$ |\text{e3}\rangle =B^2 \lambda  Q \bar{\mu } \bar{Q}$,
$|\text{e4}\rangle = |\text{e5}\rangle = |\text{e6}\rangle = |\text{e7}\rangle =|\text{e8}\rangle =0$. The associated eigenfunctions are extracted as represented below:
\begin{equation}\label{101202}
|\Theta_1\rangle=
\left(
\begin{array}{c}
 0 \\
 1 \\
 0 \\
 0 \\
 0 \\
 0 \\
 0 \\
 0 \\
\end{array}
\right), 
|\Theta_2\rangle=
\left(
\begin{array}{c}
 \frac{A}{B \sqrt{\left| \frac{A}{B \bar{\lambda }}\right| ^2+1} \bar{\lambda }} \\
 0 \\
 0 \\
 0 \\
 0 \\
 0 \\
 0 \\
 \frac{1}{\sqrt{\left| \frac{A}{B \bar{\lambda }}\right| ^2+1}} \\
\end{array}
\right)
|\Theta_3\rangle=
\left(
\begin{array}{c}
 0 \\
 0 \\
 0 \\
 1 \\
 0 \\
 0 \\
 0 \\
 0 \\
\end{array}
\right).
\end{equation}
By making use of the eigenvalues as well as the corresponding eigenfunctions in Eq. (\ref{062811}) while considering the fact that $\langle\psi_i|\frac{\partial\psi_i}{\partial \lambda}\rangle=0$, the classical and quantum portions of the QFI are gotten with regard to $\theta$, and these portions are summed up to obtain:
\begin{equation}\label{010401}
\begin{split}
sF\theta_{SGAD} &= \frac{4 (1-\lambda)^2 Q^2 e^{w/T}}{(1-\lambda)^2 \sin^2(\theta) \left(e^{w/T} + 1\right) + \cos^2(\theta) e^{w/T}} \\
&\quad + \frac{Q^2 \left( 2 (1-\lambda)^2 \sin(\theta) \cos(\theta) - \frac{2 \sin(\theta) \cos(\theta)}{e^{-w/T}+1} \right)^2}{(1-\lambda)^2 \sin^2(\theta)} \\
&\quad + \frac{\cos^2(\theta)}{e^{-w/T}+1} + 4 \lambda (1-\mu) (1-Q) Q \cos^2(\theta) \\
&\quad + \frac{Q \left( 2 \lambda \mu (1-Q) e^{2 i \Phi} \sin(\theta) \cos(\theta) + Q \left( 2 \lambda^2 \sin(\theta) \cos(\theta) - \frac{2 \sin(\theta) \cos(\theta)}{e^{w/T}+1} \right) \right)^2}{\lambda \mu (1-Q) e^{2 i \Phi} \sin^2(\theta) + Q \left( \lambda^2 \sin^2(\theta) + \frac{\cos^2(\theta)}{e^{w/T}+1} \right)}.
\end{split}
\end{equation}
Also, the QFI regarding the $\phi$ parameter is obtained as below:
\begin{equation}\label{101204}
sF\phi_{SGAD}  =\frac{4 (1-\lambda )^2 Q^2 \sin ^2(\theta ) \cos ^2(\theta ) e^{w/T}}{(1-\lambda )^2 \sin ^2(\theta ) \left(e^{w/T}+1\right)+\cos ^2(\theta ) e^{w/T}}.
\end{equation}
\begin{figure}[!t]
	\centering
		\includegraphics[width=0.5\textwidth]{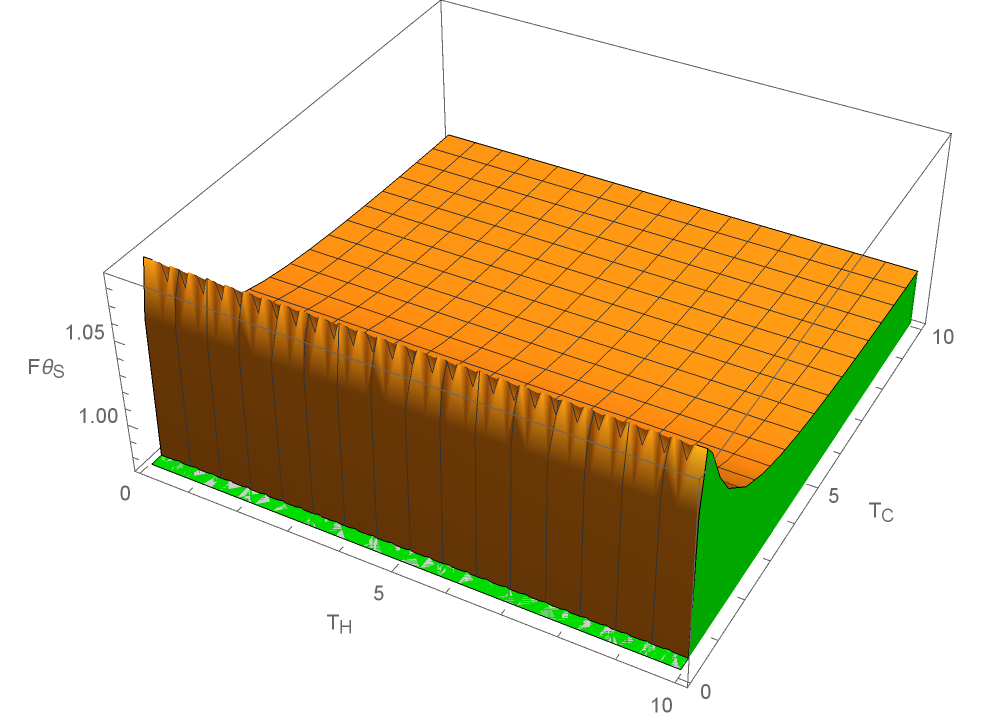}
	\caption{In the scenario when $r=1$ and $\theta=0$, the 3D graph shows variations in $F\theta_{SGAD}$ with modifications in the SGAD channel temperature $T_C$ and the Hawking temperature $T_H$.}
	\label{fig3}
\end{figure}

\begin{figure}[!t]
	\centering
		\includegraphics[width=0.75\textwidth]{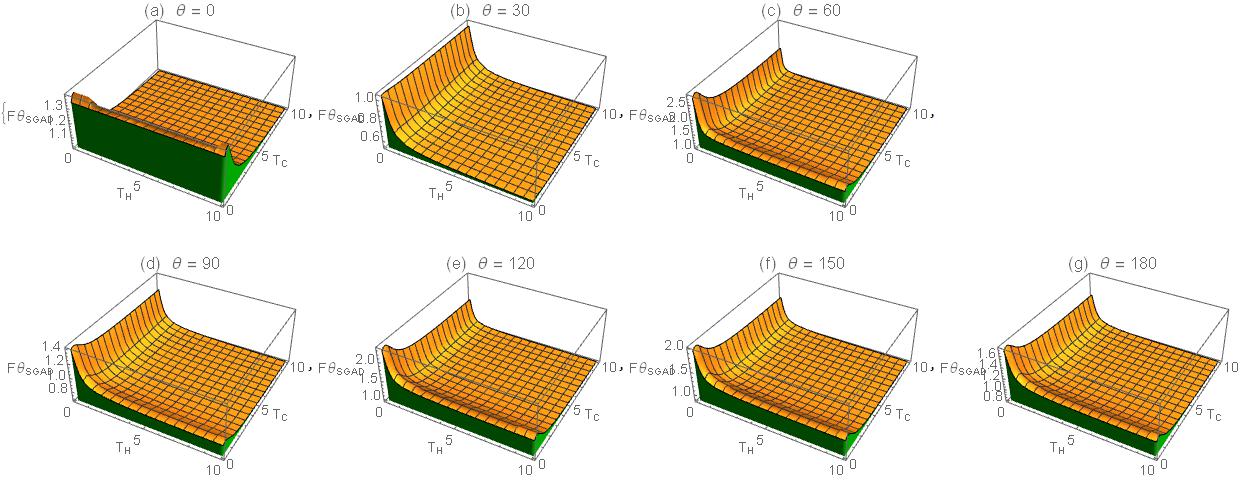}  
	\caption{ (a-g) 3D graph illustrating changes in $F\theta_{SGAD}$ associated with variations in the SGAD channel temperature $T_C$, the Hawking temperature $T_H$, and the angle $\theta$, while the extent of squeezing r is fixed at 0.}
\label{fig4}
\end{figure}

\begin{figure}[!t]
\centering
\includegraphics[width=0.5\textwidth]{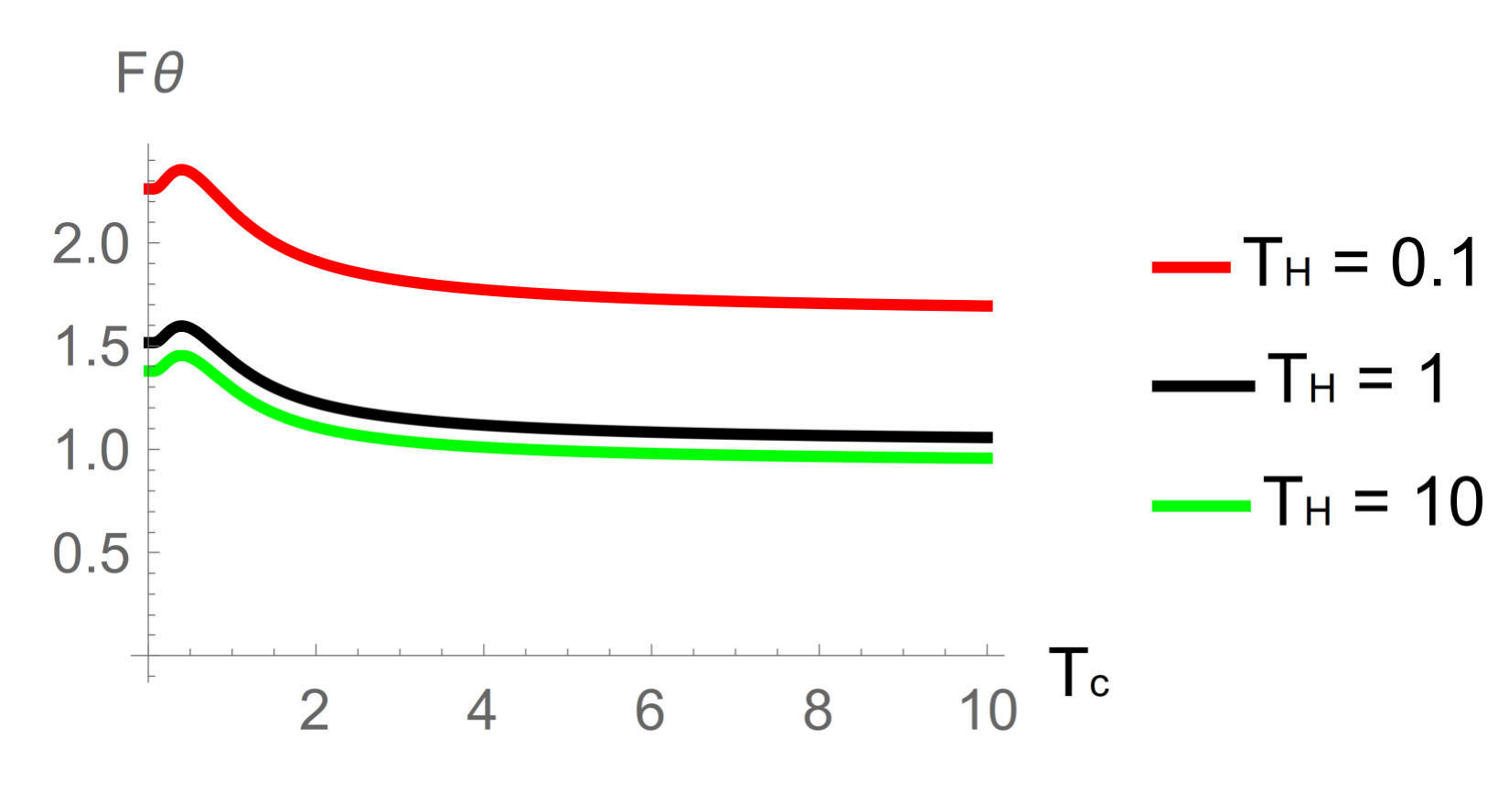} 
\caption{Graphs illustrating changes in $F\theta_{SGAD}$ associated with variations in the SGAD channel temperature $T_C$ and the Hawking temperature $T_H$.}
	\label{fig5}
\end{figure}


\begin{figure}[!t]
	\centering
		\includegraphics[width=0.5\textwidth]{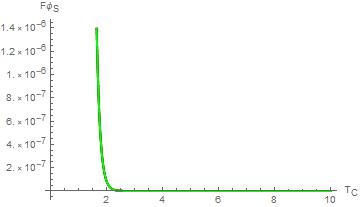}
	\caption{The relationship between the variable $sF\phi_{SGAD}$ and variations in both the temperature of the SGAD channel ($T_C$) and the Hawking temperature ($T_H$) in the presence of a Schwarzschild black hole.}
	\label{fig6}
\end{figure}
The properties of the QFI under the influence of the SGAD channel in the Schwarzschild black hole environment are illustrated in Figures~\ref{fig4}, \ref{fig5}, and \ref{fig6}. Figure~\ref{fig4} presents the variation of QFI $F_\theta^{\text{SGAD}}$ with respect to the SGAD channel temperature $T_C$ and the Hawking temperature $T_H$, under the condition where the squeezing parameter $r = 1$ and the entanglement weight parameter $\theta = 0$. In this case, $F_\theta^{\text{SGAD}}$ is observed to remain constant with increasing $T_H$, indicating robustness against Hawking radiation. However, a notable degradation in QFI is observed with increasing $T_C$, until it stabilizes at a fixed asymptotic value. It is important to note that when $r = 1$, the QFI becomes ill-defined or singular for $\theta > 0$, reflecting a sensitivity to parameter values under strong squeezing.

Figure~\ref{fig5}(a--g) explores the dependence of $F_\theta^{\text{SGAD}}$ on both $T_C$ and $T_H$ across a range of $\theta$ values. The results highlight a strong angular dependence in QFI behavior:
\begin{itemize}
  \item At $\theta = 0^\circ$, $\theta^{\text{SGAD}}$ is nearly invariant with respect to $T_H$ but decreases significantly with increasing $T_C$.
  \item At $\theta = 30^\circ$, the QFI remains mostly unaffected by $T_C$ but diminishes with increasing $T_H$.
  \item At $\theta = 60^\circ$, a slight reduction in QFI is observed with increasing $T_H$, while the influence of $T_C$ is more pronounced.
  \item For higher angles ($\theta = 90^\circ$ to $180^\circ$), the QFI continues to show greater sensitivity to $T_C$ than to $T_H$.
\end{itemize}
The figures suggest that the SGAD channel temperature $T_C$ exerts a more detrimental effect on QFI than the Hawking temperature $T_H$, indicating that environmental decoherence dominates over relativistic effects in degrading quantum information.

Figure~\ref{fig5} further illustrates the interplay between $T_C$ and $T_H$ on $F_\theta^{\text{SGAD}}$. At low values of $T_C$, a slight enhancement in QFI is observed, likely due to thermal excitation improving estimation sensitivity. However, beyond a certain threshold, QFI steadily decreases with increasing $T_C$ and eventually saturates. Additionally, higher values of $T_H$ amplify the rate of QFI degradation, reaffirming that both thermal and gravitational sources of decoherence contribute to the overall information loss, with $T_C$ being the more dominant factor.

Figure~\ref{fig6} illustrates the behavior of the QFI with respect to the parameter $\phi$ in the SGAD channel under the influence of a Schwarzschild black hole. The graph reveals that $F_\phi^{\text{SGAD}}$ exhibits a sharp and fragile spike when the channel temperature reaches $T_C = 2$. Beyond this point, the QFI decreases and remains suppressed at higher values of $T_C$. This transient spike may be attributed to any of the following phenomena: a resonance-like enhancement in phase estimation precision caused by thermal fluctuations; a sudden redistribution of quantum information resulting in a temporary boost in $F_\phi^{\text{SGAD}}$; or a non-monotonic decoherence effect in which the system momentarily regains phase sensitivity before succumbing to environmental noise. Importantly, $F_\phi^{\text{SGAD}}$ appears to be unaffected by the Hawking temperature $T_H$, as the spike remains invariant and identical to the scenario in which no black hole is present.

\subsection{Generalised amplitude damping channel}
A quantum noise framework known as the GAD channel explains how a qubit interacts with a heat reserve at a particular temperature. Unlike the standard AD channel, which assumes zero-temperature dissipation, the GAD channel accounts for thermal excitations in the environment, making it particularly relevant in quantum communication and computation scenarios.

The Kraus operators corresponding to the GAD channel are presented as below \citep{Srikanth2008squeezed}:
\begin{equation}\label{q9}
\begin{split}
E_0^{G}=&
\sqrt{Q}
	\begin{bmatrix} 
	1 & 0  \\
	0 & \sqrt{1-\lambda_{G}}
	\end{bmatrix},\\
	\quad	
E_1^{G}=&
\sqrt{Q}
	\begin{bmatrix} 
	0 & \sqrt{\lambda_{G}}  \\
	0 & 0
	\end{bmatrix},\\
	\quad
	E_2^{G}=&
	\sqrt{1-Q}
	\begin{bmatrix} 
	\sqrt{1-\lambda_{G}} & 0  \\
	0 & 1
	\end{bmatrix},\\
	\quad
	E_3^{G}=&
	\sqrt{1-Q}
	\begin{bmatrix} 
	0 & 0  \\
	\sqrt{\lambda_{G}} & 0 
	\end{bmatrix}.\\
	\quad
	\end{split}
\end{equation}
From the SGAD Kraus operators in Eq. (\ref{q6}), we obtain the GAD channel Kraus operators by substituting $\Phi$ with 0, $\mu$ with 0, and $v$ with $\lambda$.  The following equation can be used to mathematically characterize how the GAD channel affects the sent data’s quantum state:
\begin{equation}\label{q9a}
\xi(\rho)_G=\sum_{i,j}{(E_i^{G1}\otimes E_i^{G4}\otimes E_j^{G2}\otimes E_j^{G5})\rho (E_i^{G1}\otimes E_i^{G4}\otimes E_j^{G2} \otimes E_j^{G5})^\dagger}.
\end{equation}
The GAD channel is essential for understanding decoherence in quantum systems and for researching how thermal noise affects quantum data processing activities.

\subsubsection{QFI in GAD channel in the background of a Schwarzschild Black Hole}
The total QFI for the GAD channel, which includes both the classical and quantum components, with regard to $\theta$ is obtained by following similar procedure as before:
\begin{equation}\label{092903}
\begin{cases}
sF\theta_{GAD} = &\frac{Q^2 \left( 2 \lambda^2 \sin(\theta) \cos(\theta) - \frac{2 \sin(\theta) \cos(\theta)}{e^{w/T}+1} \right)^2}{\lambda^2 \sin^2(\theta) + \frac{\cos^2(\theta)}{e^{w/T}+1}} \\
&+ \frac{4 (1-\lambda)^2 Q^2 e^{w/T}}{(1-\lambda)^2 \sin^2(\theta) \left( e^{w/T} + 1 \right) + \cos^2(\theta) e^{w/T}} \\
&+ \frac{Q^2 \left( 2 (1-\lambda)^2 \sin(\theta) \cos(\theta) - \frac{2 \sin(\theta) \cos(\theta)}{e^{-w/T}+1} \right)^2}{(1-\lambda)^2 \sin^2(\theta) + \frac{\cos^2(\theta)}{e^{-w/T}+1}} \\
&+ 4 \lambda (1-Q) Q \cos^2(\theta).
\end{cases}
\end{equation}
The QFI with regard to the variable $\phi$ is also obtained and represented by:
\begin{equation}\label{092904}
sF\phi_{GAD} =\frac{4 (1-\lambda )^2 Q^2 \sin ^2(\theta ) \cos ^2(\theta ) e^{w/T}}{(1-\lambda )^2 \sin ^2(\theta ) \left(e^{w/T}+1\right)+\cos ^2(\theta ) e^{w/T}}.
\end{equation}

\begin{figure}[!t]
	\centering	
	\includegraphics[scale=1]{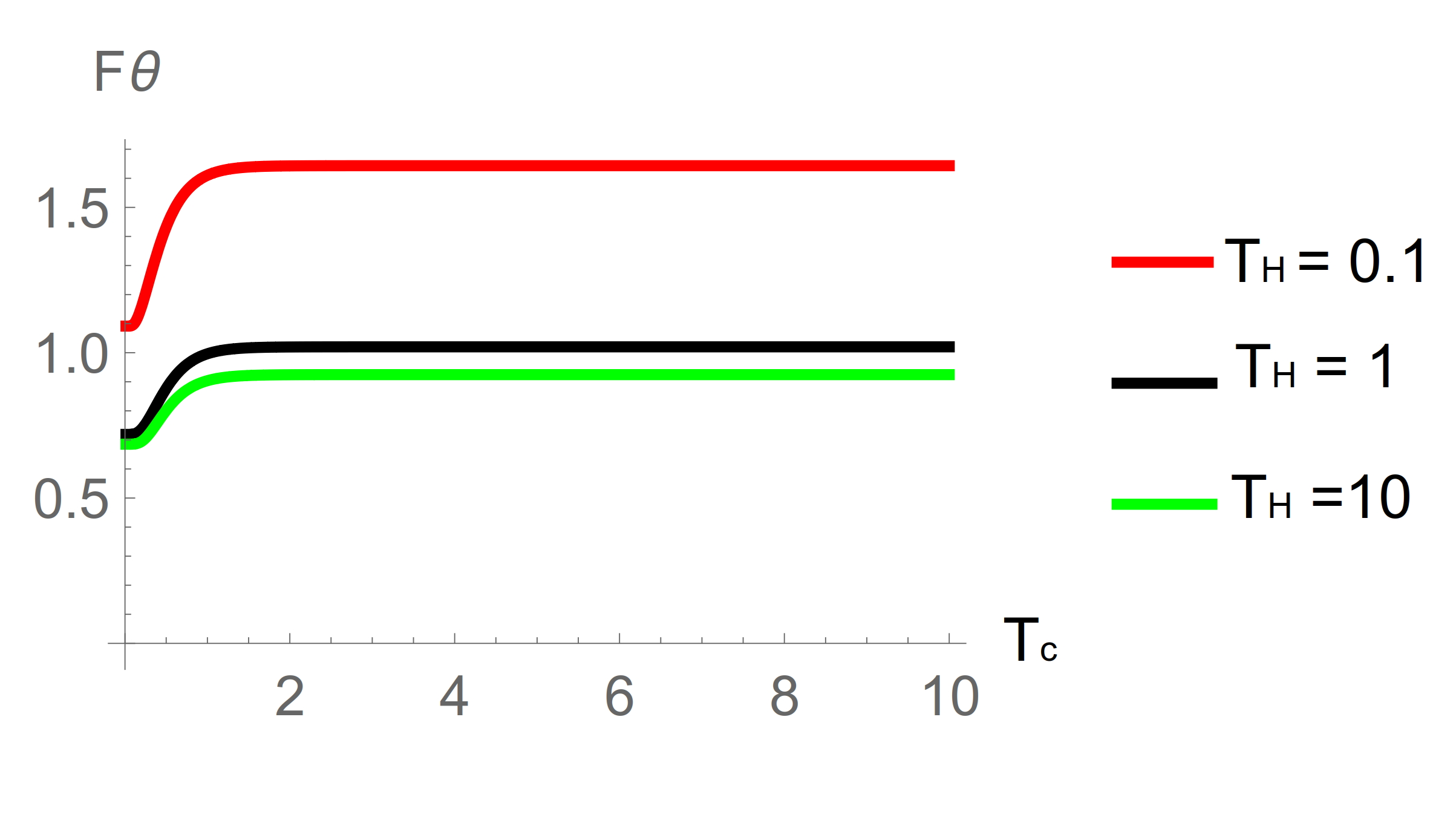}  

	\caption{Figure illustrating the relationship between the GAD channel temperature ($T_C$), the Hawking temperature ($T_H$), and the ﬂuctuation of $F_\theta^{\text{GAD}}$.}
	\label{fig7}
\end{figure}

\begin{figure}[!t]
	\centering	
	\includegraphics[width=0.5\textwidth]{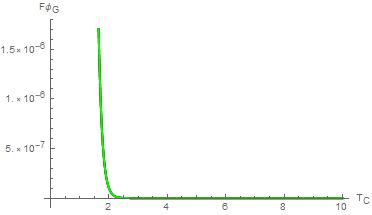} 

	\caption{The relationship between the GAD channel temperature ($T_C$), the Hawking temperature ($T_H$), and the fluctuation of $F\phi_{GAD}$.}
	\label{fig8}
\end{figure}
Figures~\ref{fig8} and~\ref{fig7} depict the behavior of the QFI in the GAD channel within the vicinity of a Schwarzschild black hole. Figure~\ref{fig8} illustrates the variation of the QFI with respect to the phase parameter $\phi$, denoted as $F_\phi^{\text{GAD}}$, as both the channel temperature $T_C$ and the Hawking temperature $T_H$ increase. A prominent and sharp spike is observed in $F_\phi^{\text{GAD}}$ when the channel temperature reaches approximately $T_C = 2$. Notably, this feature appears invariant under changes in the Hawking temperature, as all curves corresponding to different values of $T_H$ overlap. This suggests that the QFI with respect to $\phi$ is highly sensitive to a specific thermal condition in the GAD environment, potentially indicating a transient resonance or phase coherence revival. However, beyond this temperature, the QFI diminishes, implying that decoherence overtakes any transient amplification. The independence from $T_H$ implies that the curved spacetime effect specifically the Hawking radiation does not significantly influence phase estimation in the GAD context, reinforcing the dominance of environmental (thermal) decoherence over relativistic effects in this regime.

Figure~\ref{fig7} presents the variation of QFI with respect to the weight parameter $\theta$, denoted $F_\theta^{\text{GAD}}$, under increasing $T_C$ and $T_H$. The figure shows that $F_\theta^{\text{GAD}}$ improves as $T_C$ increases, eventually approaching a saturation point. This enhancement can be attributed to thermal excitation that initially boosts sensitivity to changes in $\theta$. Nevertheless, as the Hawking temperature $T_H$ increases, the QFI progressively declines, indicating that gravitational decoherence due to Hawking radiation becomes more detrimental to entanglement and parameter estimation. The trade-off between the channel-induced excitation and the relativistic loss of coherence highlights a complex interplay in which thermal noise can both help and hinder QFI, depending on the temperature regime. This behavior has direct implications for quantum communication protocols such as QT and QKD, which rely heavily on the preservation of entanglement fidelity.

Figures~\ref{fig8} and~\ref{fig7} reinforce the observation that, in the GAD channel, $T_C$ acts as both a facilitator and a detractor depending on the parameter and thermal regime, while $T_H$ predominantly contributes to information loss through decoherence.

\subsection{Amplitude Damping channel}
An essential quantum noise model for describing energy dissipation in a quantum system is the AD channel. It is especially important when a quantum system contacts with its surroundings and causes a permanent depletion in excitation. In quantum optics, this kind of noise is frequently used to simulate the spontaneous emission of photons from an excited atom or qubit. The AD channel's Kraus operators are displayed below \citep{Srikanth2008squeezed}: 
\begin{equation}\label{q7}
E_0^{A}=
	\begin{bmatrix} 
	1 & 0  \\
	0 & \sqrt{1-\lambda_{A}}\\
	\end{bmatrix},
	\quad
E_1^{A}=
	\begin{bmatrix} 
	0 & \sqrt{\lambda_A}  \\
	0 & 0\\
	\end{bmatrix}.
	\quad
\end{equation}
When particles move over an AD channel, the rate of decoherence $\lambda_A$ (where $0 \leq \lambda \leq 1$) measures the likelihood of error. Understanding the AD channel is crucial for quantum communication and computation, as it helps in designing error correction schemes to mitigate decoherence effects.

\subsubsection{QFI in AD channel in the background of a Schwarzschild Black Hole}
\begin{figure}[!t]	\centering
		\includegraphics[width=0.5\textwidth]{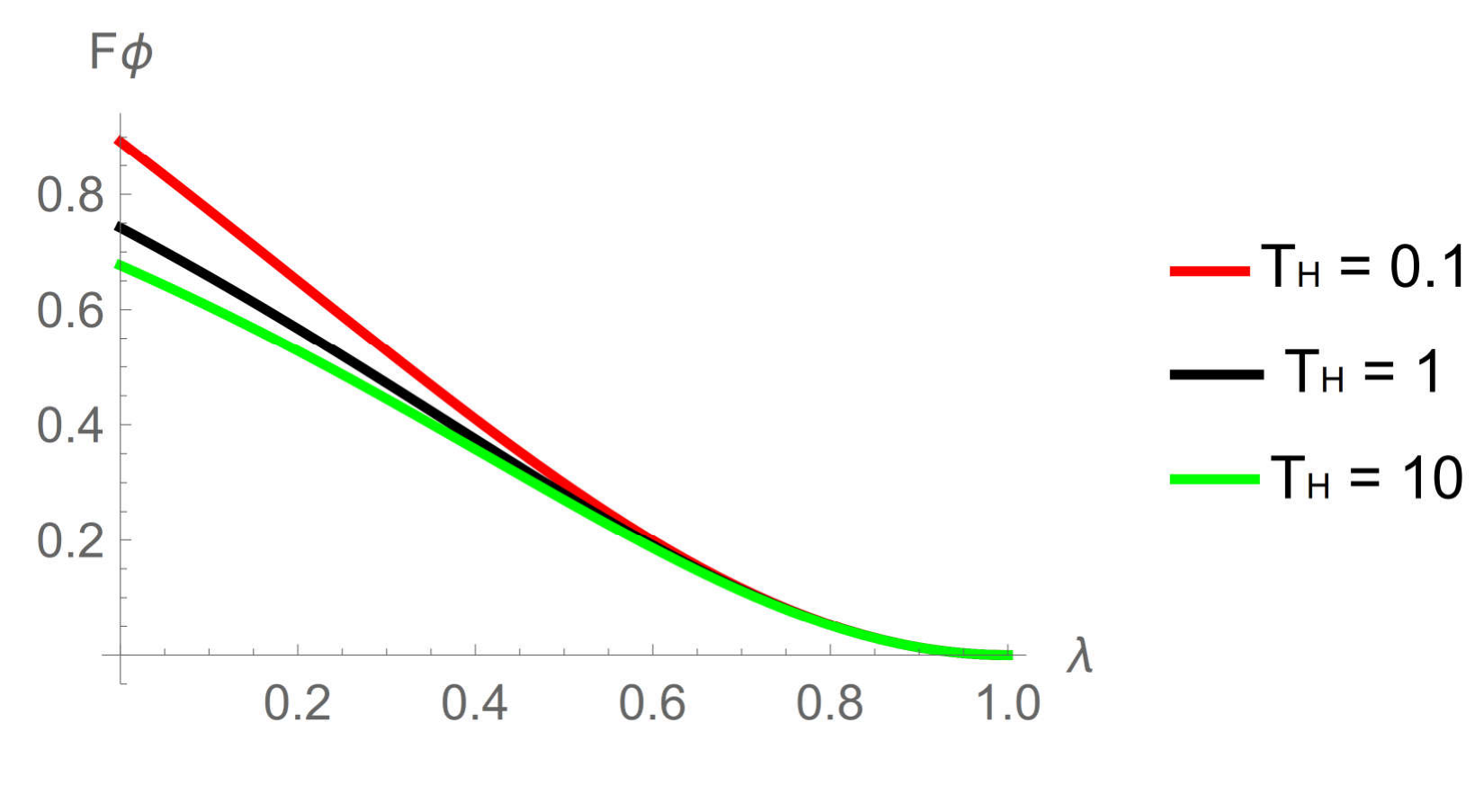}
	\caption{ Graph showing the correlation between changes regarding the AD channel variable $\lambda$ and the value of $F\phi_{AD}$.}
	\label{fig9}
\end{figure}

\begin{figure}[!t]
	\centering
		\includegraphics[width=0.5\textwidth]{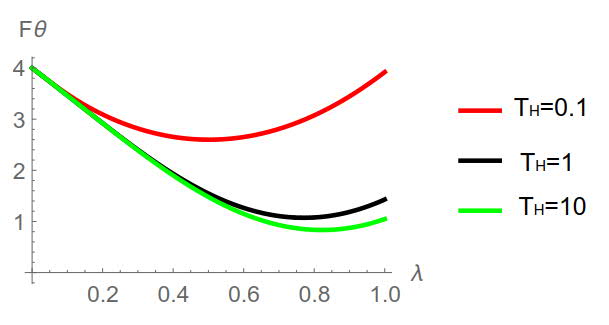}
	\caption{ Graph illustrating the relationship between the value of $F\theta_{AD}$ and variations in the AD noise variable $\lambda$.}
	\label{fig:stsAD1}
\end{figure}

The entire QFI in relation to the variable $\theta$ may be expressed using the same procedure outlined in previous sections:
\begin{equation}\label{092901}
\begin{split}
|\text{sF}\theta_{\text{AD}}\rangle = & \frac{4 \bar{\lambda}^2 e^{w/T}}{\bar{\lambda}^2 \sin^2(\theta) \left( e^{w/T} + 1 \right) + \cos^2(\theta) e^{w/T}} \\
& + \frac{\left( 2 \bar{\lambda}^2 \sin(\theta) \cos(\theta) - \frac{2 \sin(\theta) \cos(\theta)}{e^{-w/T} + 1} \right)^2}{\bar{\lambda}^2 \sin^2(\theta) + \frac{\cos^2(\theta)}{e^{-w/T} + 1}} \\
& + \frac{\left( 2 \lambda^2 \sin(\theta) \cos(\theta) - \frac{2 \sin(\theta) \cos(\theta)}{e^{w/T} + 1} \right)^2}{\lambda^2 \sin^2(\theta) + \frac{\cos^2(\theta)}{e^{w/T} + 1}}.
\end{split}
\end{equation}
The overall QFI with regard to $\phi$ is represented by:
\begin{equation}\label{092902}
sF\phi_{AD}=\frac{(\lambda -1)^2 \sin ^2(2 \theta ) e^{w/T}}{(\lambda -1)^2 \sin ^2(\theta ) \left(e^{w/T}+1\right)+\cos ^2(\theta ) e^{w/T}}.
\end{equation}
Here, $\lambda$ represents the variable for amplitude-damping noise, and $T_H$ represents the Hawking temperature, which arises from the Schwarzschild Black Hole. In Figures (\ref{fig9}) and (\ref{fig:stsAD1}), The Schwarzschild Black Hole background's QFI characteristics in the AD channel are displayed in figure (\ref{fig9}). 
Figure (\ref{fig9}) depicts the relationship between $F\phi_{AD}$ and the increase in the AD channel noise variable and the Hawking temperature $T_H$. The decoherence of QFI in relation to $\phi$ in the AD channel $F\phi_{AD}$ is shown to rise with increasing values of the AD noise variable $\lambda$ and Hawking temperature $T_H$. The image illustrates that at greater values of the AD noise parameter $\lambda_{AD}$, the QFI regarding $\phi$ becomes unaffected by the Hawking temperature $T_H$.
Figure (\ref{fig:stsAD1}) illustrates the relationship between $F\theta_{AD}$ and the AD channel noise parameter $\lambda$. It demonstrates that, in general, $F\theta_{AD}$ first experiences a decrease in coherence as $\lambda$ increases, but then recovers and becomes stronger at higher values of $\lambda$. The magnitude of its reduction is contingent upon the Hawking temperature $T_H$. The occurrence of decoherence in $F\theta_{AD}$ is evident as the Hawking temperature $T_H$ increases. The graphic suggests that $F\theta_{AD}$ may return to its starting value at lower Hawking temperatures $T_H$ compared to higher Hawking temperatures $T_H$. Furthermore, since all the charts match at the origin of the graph, it is clear that the QFI $\theta$ is unaffected by the Hawking temperature $T_H$ for very low values of the noise parameter $\lambda$.

\section{Discussion and Conclusion} \label{five}
This study has investigated the QFI of a Dirac system interacting with dissipative noisy channels in the curved spacetime of a Schwarzschild black hole, with particular focus on the influence of Hawking radiation (characterized by the Hawking temperature $T_H$) and environmental decoherence (captured by the channel temperature $T_C$). The behavior of the QFI with respect to both entanglement weight ($\theta$) and phase ($\phi$) parameters was analyzed for three distinct noise models: the SGAD channel and its subchannels -- GAD and AD.

In the SGAD channel, the squeezing angle $\Phi$ was found to have no effect on the QFI with respect to $\theta$. However, when the squeezing parameter was set to $r = 1$, the QFI $F_\theta$ became completely robust against variations in the Hawking temperature $T_H$, demonstrating resistance to relativistic decoherence. Despite this, $F_\theta$ still exhibited sensitivity to the SGAD channel temperature $T_C$, with higher values leading to information degradation. Notably, the magnitude of QFI decay with respect to $T_C$ was significantly smaller for $r = 1$ than for $r = 0$, suggesting that squeezing serves as a potential error-correction mechanism.

The GAD channel analysis revealed that an increase in $T_C$ initially enhances $F_\theta$ before it plateaus, whereas $T_H$ induces decoherence and suppresses QFI. This indicates that thermal excitation can boost parameter sensitivity up to a threshold, beyond which noise dominates. A singular spike in $F_\phi$ was observed around $T_C = 2$, irrespective of $T_H$, possibly linked to thermal resonance or a non-monotonic revival in phase sensitivity.

In the AD channel, $F_\theta$ displayed a non-trivial dependence on the noise parameter $\lambda$: at low $T_H$, the QFI was observed to recover with increasing $\lambda$, indicating that under minimal gravitational decoherence, noise-induced excitation can restore information. However, at higher $T_H$, QFI remained suppressed, confirming that Hawking radiation disrupts coherence more significantly under strong gravitational influence.

Overall, the results indicate that:
\begin{itemize}
    \item The SGAD channel demonstrates the most potential for protecting QFI via squeezing.
    \item The GAD channel highlights a trade-off between thermal excitation and relativistic decoherence.
    \item The AD channel reveals that decoherence can be partially reversible under specific thermal-gravitational regimes.
\end{itemize}
It is thus concluded that while the Hawking temperature $T_H$ does contribute to decoherence and information loss, its effects are generally less severe than those of the environmental channel temperature $T_C$. Moreover, squeezing ($r=1$) provides a promising strategy for mitigating quantum information loss, acting as a form of quantum error correction in curved spacetime. These insights are significant for the development of resilient quantum communication protocols in relativistic and noisy environments. The methodology used in this research and the results are in agreement with other similar works carried out by other authors \citep{huang2018protecting,huang2020quantum}

\section*{Declarations}
\subsection*{Availability of data and materials}
All data generated or analysed during this study are contained within this published manuscript. No additional external datasets were used or are required.

\subsection*{Competing interests}
The authors declare that they have no competing interests.

\subsection*{Author Contributions}
C.I. conceptualized the research idea, developed the theoretical framework, and performed the analytical derivations. M.S.L. contributed to the formulation of the mathematical models of the study. B.O.A. assisted in numerical simulations, data visualization, and preparation of the figures. E.S.J. was responsible for the literature review, reference management, and assisted in the manuscript editing. Y.M.P. contributed to figure formatting, and assisted in result interpretation. B.J.F. provided overall project supervision, validated the scientific content, and contributed to the final writing, structuring, and critical revision of the manuscript. All authors reviewed and approved the final version of the manuscript.

\section*{Funding}
This research was supported by the Tertiary Education Trust Fund (TETFund) and Federal University Lafia (FULafia) under the Institution Based Research (IBR) program. Project Code: FUL/REG/TETFund/002/VOL.V1/329.
\bibliographystyle{unsrt}
\bibliography{QFI_2}

\begin{thebibliography}{10}

\bibitem{krasnoholovets2022information}
Volodymyr Krasnoholovets.
\newblock Information field and its carriers in biological systems.
\newblock {\em NeuroQuantology}, 20(4):179--201, 2022.

\bibitem{mishra2022key}
Aman~Kumar Mishra and Vijayakumar Ponnusamy.
\newblock Key technologies and architectures for 6g and beyond wireless
  communication system.
\newblock In {\em AI and Blockchain Technology in 6G Wireless Network}, pages
  25--44. Springer, 2022.

\bibitem{azahari2023quantum}
Nur Shahirah~Binti Azahari, Nur Ziadah~Binti Harun, and Zuriati Binti~Ahmad
  Zukarnain.
\newblock Quantum identity authentication for non-entanglement multiparty
  communication: A review, state of art and future directions.
\newblock {\em ICT Express}, 9(4):534--547, 2023.

\bibitem{iyen2023scrutinizing}
Cookey Iyen, Babatunde~James Falaye, and Muhammad~Sanusi Liman.
\newblock Scrutinizing joint remote state preparation under decoherence.
\newblock {\em Scientific Reports}, 13(1):8066, 2023.

\bibitem{einstein1935can}
Albert Einstein, Boris Podolsky, and Nathan Rosen.
\newblock Can quantum-mechanical description of physical reality be considered
  complete?
\newblock {\em Physical review}, 47(10):777, 1935.

\bibitem{zygelman2024no}
Bernard Zygelman.
\newblock No-cloning theorem, quantum teleportation and spooky correlations.
\newblock In {\em A First Introduction to Quantum Computing and Information},
  pages 123--145. Springer, 2024.

\bibitem{venkatesh2024lightweight}
Ranjitha Venkatesh.
\newblock A lightweight quantum blockchain-based framework to protect patients
  private medical information.
\newblock {\em IEEE Transactions on Network Science and Engineering}, 2024.

\bibitem{castro2022enhancing}
GS~Castro and Rubens~Viana Ramos.
\newblock Enhancing eavesdropping detection in quantum key distribution using
  disentropy measure of randomness.
\newblock {\em Quantum Information Processing}, 21(2):79, 2022.

\bibitem{lee2022eavesdropping}
Chankyun Lee, Ilkwon Sohn, and Wonhyuk Lee.
\newblock Eavesdropping detection in bb84 quantum key distribution protocols.
\newblock {\em IEEE Transactions on Network and Service Management},
  19(3):2689--2701, 2022.

\bibitem{bakhshinezhad2024scalable}
Pharnam Bakhshinezhad, Mohammad Mehboudi, Carles Roch~I Carceller, and Armin
  Tavakoli.
\newblock Scalable entanglement certification via quantum communication.
\newblock {\em PRX Quantum}, 5(2):020319, 2024.

\bibitem{uppu2021quantum}
Ravitej Uppu, Leonardo Midolo, Xiaoyan Zhou, Jacques Carolan, and Peter Lodahl.
\newblock Quantum-dot-based deterministic photon--emitter interfaces for
  scalable photonic quantum technology.
\newblock {\em Nature nanotechnology}, 16(12):1308--1317, 2021.

\bibitem{khan2024quantum}
Muhammad~Annas Khan, Salman Ghafoor, Syed Mohammad~Hassan Zaidi, Haibat Khan,
  and Arsalan Ahmad.
\newblock From quantum communication fundamentals to decoherence mitigation
  strategies: Addressing global quantum network challenges and projected
  applications.
\newblock {\em Heliyon}, 2024.

\bibitem{martinez2022decoherence}
Josu~Etxezarreta Martinez.
\newblock Decoherence and quantum error correction for quantum computing and
  communications.
\newblock {\em arXiv preprint arXiv:2202.08600}, 2022.

\bibitem{hu2024quantum}
Le~Hu.
\newblock {\em Quantum Dynamics: Quantum Metrology, Control, Open Systems and
  Entanglement Dynamics}.
\newblock University of Rochester, 2024.

\bibitem{zhou2025randomized}
Sisi Zhou and Senrui Chen.
\newblock Randomized measurements for multi-parameter quantum metrology.
\newblock {\em arXiv preprint arXiv:2502.03536}, 2025.

\bibitem{jin2024quantum}
Rui-Bo Jin, Zi-Qi Zeng, Chenglong You, and Chenzhi Yuan.
\newblock Quantum interferometers: principles and applications.
\newblock {\em Progress in Quantum Electronics}, page 100519, 2024.

\bibitem{marshall2022high}
Mason~C Marshall, Reza Ebadi, Connor Hart, Matthew~J Turner, Mark~JH Ku,
  David~F Phillips, and Ronald~L Walsworth.
\newblock High-precision mapping of diamond crystal strain using quantum
  interferometry.
\newblock {\em Physical Review Applied}, 17(2):024041, 2022.

\bibitem{sinatra2022spin}
Alice Sinatra.
\newblock Spin-squeezed states for metrology.
\newblock {\em Applied Physics Letters}, 120(12), 2022.

\bibitem{li2023multi}
Yue Li, Xu~Cheng, Lingna Wang, Xingyu Zhao, Waner Hou, Yi~Li, Kamran Rehan,
  Mingdong Zhu, Lin Yan, Xi~Qin, et~al.
\newblock Multi-parameter quantum metrology with stabilized multi-mode squeezed
  state.
\newblock {\em arXiv preprint arXiv:2312.10379}, 2023.

\bibitem{huang2024entanglement}
Jiahao Huang, Min Zhuang, and Chaohong Lee.
\newblock Entanglement-enhanced quantum metrology: from standard quantum limit
  to heisenberg limit.
\newblock {\em Applied Physics Reviews}, 11(3), 2024.

\bibitem{long2022entanglement}
Xinyue Long, Wan-Ting He, Na-Na Zhang, Kai Tang, Zidong Lin, Hongfeng Liu,
  Xinfang Nie, Guanru Feng, Jun Li, Tao Xin, et~al.
\newblock Entanglement-enhanced quantum metrology in colored noise by quantum
  zeno effect.
\newblock {\em Physical Review Letters}, 129(7):070502, 2022.

\bibitem{deng2024quantum}
Xiaowei Deng, Sai Li, Zi-Jie Chen, Zhongchu Ni, Yanyan Cai, Jiasheng Mai, Libo
  Zhang, Pan Zheng, Haifeng Yu, Chang-Ling Zou, et~al.
\newblock Quantum-enhanced metrology with large fock states.
\newblock {\em Nature Physics}, pages 1--7, 2024.

\bibitem{altherr2021quantum}
Anian Altherr and Yuxiang Yang.
\newblock Quantum metrology for non-markovian processes.
\newblock {\em Physical Review Letters}, 127(6):060501, 2021.

\bibitem{rath2021quantum}
Aniket Rath, Cyril Branciard, Anna Minguzzi, and Beno{\^\i}t Vermersch.
\newblock Quantum fisher information from randomized measurements.
\newblock {\em Physical Review Letters}, 127(26):260501, 2021.

\bibitem{gorecki2022quantum}
Wojciech G{\'o}recki, Alberto Riccardi, and Lorenzo Maccone.
\newblock Quantum metrology of noisy spreading channels.
\newblock {\em Physical Review Letters}, 129(24):240503, 2022.

\bibitem{das2025investigating}
Subrata Das, Avimita Chatterjee, and Swaroop Ghosh.
\newblock Investigating impact of bit-flip errors in control electronics on
  quantum computation.
\newblock In {\em 2025 38th International Conference on VLSI Design and 2024
  23rd International Conference on Embedded Systems (VLSID)}, pages 558--563.
  IEEE, 2025.

\bibitem{hu2020protecting}
Ming-Liang Hu and Hui-Fang Wang.
\newblock Protecting quantum fisher information in correlated quantum channels.
\newblock {\em Annalen der Physik}, 532(1):1900378, 2020.

\bibitem{khatri2020information}
Sumeet Khatri, Kunal Sharma, and Mark~M Wilde.
\newblock Information-theoretic aspects of the generalized amplitude-damping
  channel.
\newblock {\em Physical Review A}, 102(1):012401, 2020.

\bibitem{zlotnick2025entanglement}
Elyakim Zlotnick, Boulat Bash, and Uzi Pereg.
\newblock Entanglement-assisted covert communication via qubit depolarizing
  channels.
\newblock {\em IEEE Transactions on Information Theory}, 2025.

\bibitem{rahman2022fidelity}
Atta~Ur Rahman, Saeed Haddadi, Mohammad~Reza Pourkarimi, and Mehrdad
  Ghominejad.
\newblock Fidelity of quantum states in a correlated dephasing channel.
\newblock {\em Laser Physics Letters}, 19(3):035204, 2022.

\bibitem{chessa2021quantum}
Stefano Chessa and Vittorio Giovannetti.
\newblock Quantum capacity analysis of multi-level amplitude damping channels.
\newblock {\em Communications Physics}, 4(1):22, 2021.

\bibitem{iyen2024examining}
C~Iyen, MS~Liman, SJ~Emem-Obong, WA~Yahya, CA~Onate, and BJ~Falaye.
\newblock Examining the quantum fisher information in the interaction of a
  dirac system with a squeezed generalized amplitude damping channel.
\newblock {\em Scientific Reports}, 14(1):24495, 2024.

\bibitem{han2024mass}
Hyewon Han and Bogeun Gwak.
\newblock Mass fluctuations in non-rotating btz black holes.
\newblock {\em Physics Letters B}, 857:138980, 2024.

\bibitem{kunstatter2022general}
Gabor Kunstatter and Saurya Das.
\newblock General relativity.
\newblock In {\em A First Course on Symmetry, Special Relativity and Quantum
  Mechanics: The Foundations of Physics}, pages 139--162. Springer, 2022.

\bibitem{zhao2023trapped}
Peng Zhao, David Hilditch, and Juan A~Valiente Kroon.
\newblock Trapped surface formation for the einstein-scalar system.
\newblock {\em arXiv preprint arXiv:2304.01695}, 2023.

\bibitem{wu2024genuinely}
Shu-Min Wu, Xiao-Wei Teng, Jin-Xuan Li, Si-Han Li, Tong-Hua Liu, and Jie-Ci
  Wang.
\newblock Genuinely accessible and inaccessible entanglement in schwarzschild
  black hole.
\newblock {\em Physics Letters B}, 848:138334, 2024.

\bibitem{alonso2022effective}
Asier Alonso-Bardaji, David Brizuela, and Ra{\"u}l Vera.
\newblock An effective model for the quantum schwarzschild black hole.
\newblock {\em Physics Letters B}, 829:137075, 2022.

\bibitem{horowitz2023extremal}
Gary~T Horowitz, Maciej Kolanowski, Grant~N Remmen, and Jorge~E Santos.
\newblock Extremal kerr black holes as amplifiers of new physics.
\newblock {\em Physical Review Letters}, 131(9):091402, 2023.

\bibitem{cangemi2023kerr}
Lucile Cangemi, Marco Chiodaroli, Henrik Johansson, Alexander Ochirov, Paolo
  Pichini, and Evgeny Skvortsov.
\newblock Kerr black holes from massive higher-spin gauge symmetry.
\newblock {\em Physical Review Letters}, 131(22):221401, 2023.

\bibitem{senjaya2024exact}
David Senjaya.
\newblock Exact analytical quasibound states of a scalar particle around a
  reissner-nordstr{\"o}m black hole.
\newblock {\em Physics Letters B}, 848:138373, 2024.

\bibitem{shaymatov2021charged}
Sanjar Shaymatov, Bakhtiyor Narzilloev, Ahmadjon Abdujabbarov, and Cosimo
  Bambi.
\newblock Charged particle motion around a magnetized reissner-nordstr{\"o}m
  black hole.
\newblock {\em Physical Review D}, 103(12):124066, 2021.

\bibitem{foo2021hawking}
Joshua Foo and Michael~RR Good.
\newblock Hawking radiation particle spectrum of a kerr-newman black hole.
\newblock {\em Journal of Cosmology and Astroparticle Physics}, 2021(01):019,
  2021.

\bibitem{sun2021entanglement}
Po-Chun Sun.
\newblock Entanglement islands from holographic thermalization of rotating
  charged black hole.
\newblock {\em arXiv preprint arXiv:2108.12557}, 2021.

\bibitem{harraz2021protected}
Sajede Harraz, Shuang Cong, and Juan~J Nieto.
\newblock Protected quantum teleportation through noisy channel by weak
  measurement and environment-assisted measurement.
\newblock {\em IEEE Communications Letters}, 26(3):528--531, 2021.

\bibitem{zidan2023quantum}
Nour Zidan, Atta ur~Rahman, and Saeed Haddadi.
\newblock Quantum teleportation in a two-superconducting qubit system under
  dephasing noisy channel: role of josephson and mutual coupling energies.
\newblock {\em Laser Physics Letters}, 20(2):025204, 2023.

\bibitem{mafu2022security}
Mhlambululi Mafu, Comfort Sekga, and Makhamisa Senekane.
\newblock Security of bennett--brassard 1984 quantum-key distribution under a
  collective-rotation noise channel.
\newblock In {\em Photonics}, volume~9, page 941. MDPI, 2022.

\bibitem{shu2023quantum}
Hao Shu, Chang-Yue Zhang, Yue-Qiu Chen, Zhu-Jun Zheng, and Shao-Ming Fei.
\newblock Quantum key distribution over noisy channels by the testing state
  method.
\newblock {\em International Journal of Theoretical Physics}, 62(8):160, 2023.

\bibitem{falaye2017investigating}
BJ~Falaye, AG~Adepoju, AS~Aliyu, MM~Melchor, MS~Liman, OJ~Oluwadare,
  MD~Gonz{\'a}lez-Ram{\'\i}rez, and KJ~Oyewumi.
\newblock Investigating quantum metrology in noisy channels.
\newblock {\em Scientific Reports}, 7(1):16622, 2017.

\bibitem{adepoju2017joint}
Adenike~Grace Adepoju, Babatunde~James Falaye, Guo-Hua Sun, Oscar
  Camacho-Nieto, and Shi-Hai Dong.
\newblock Joint remote state preparation (jrsp) of two-qubit equatorial state
  in quantum noisy channels.
\newblock {\em Physics Letters A}, 381(6):581--587, 2017.

\bibitem{oh2021quantum}
Changhun Oh, Sisi Zhou, Yat Wong, and Liang Jiang.
\newblock Quantum limits of superresolution in a noisy environment.
\newblock {\em Physical Review Letters}, 126(12):120502, 2021.

\bibitem{zhai2023control}
Yue Zhai, Xiaodong Yang, Kai Tang, Xinyue Long, Xinfang Nie, Tao Xin, Dawei Lu,
  and Jun Li.
\newblock Control-enhanced quantum metrology under markovian noise.
\newblock {\em Physical Review A}, 107(2):022602, 2023.

\bibitem{ragazzi2024generalized}
Giovanni Ragazzi, Simone Cavazzoni, Paolo Bordone, and Matteo~GA Paris.
\newblock Generalized phase estimation in noisy quantum gates.
\newblock {\em Physical Review A}, 110(5):052425, 2024.

\bibitem{falaye2020probing}
Babatunde~James Falaye and Muhammad~Sanusi Liman.
\newblock Probing quantum fisher information of an open dirac system with
  hawking effect in the schwarzschild black hole.
\newblock {\em Laser Physics}, 30(11):115206, 2020.

\bibitem{damour1976black}
Thibaut Damour and Remo Ruffini.
\newblock Black-hole evaporation in the klein-sauter-heisenberg-euler
  formalism.
\newblock {\em Physical Review D}, 14(2):332, 1976.

\bibitem{ghosh2012surface}
Joydip Ghosh, Austin~G Fowler, and Michael~R Geller.
\newblock Surface code with decoherence: An analysis of three superconducting
  architectures.
\newblock {\em Physical Review A}, 86(6):062318, 2012.

\bibitem{Nielsen20001}
M~Nielsen.
\newblock 1. chuang, quantum computation and quantum information. cambridge,
  uk, 2000.

\bibitem{Srikanth2008squeezed}
R~Srikanth and Subhashish Banerjee.
\newblock Squeezed generalized amplitude damping channel.
\newblock {\em Physical Review A}, 77(1):012318, 2008.

\bibitem{huang2018protecting}
Zhiming Huang.
\newblock Protecting quantum fisher information in curved space-time.
\newblock {\em The European Physical Journal Plus}, 133:1--8, 2018.

\bibitem{huang2020quantum}
Zhiming Huang, Haozhen Situ, and Zhimin He.
\newblock Quantum fisher information in the cosmic string spacetime.
\newblock {\em Classical and Quantum Gravity}, 37(17):175002, 2020.

\end{thebibliography}
\end{document}